\documentclass[11pt,a4paper]{article}
\pdfoutput=1

\usepackage[]{Packages}

\RequirePackage{placeins}
\RequirePackage{ragged2e}

\usepackage{Definitions}

\allowdisplaybreaks[1]

\preprint{\textsc{ipmu}11-0128\\ \textsc{nsf-kitp}-11-189}

\newcommand{\OfficialTitle}{General Omega Deformations\\ from Closed String Backgrounds}

\title{\vspace{2cm}
  {\huge   \textbf{\OfficialTitle}}
}

\hypersetup{pdfauthor={},pdftitle={\OfficialTitle}}

\author{
  \begin{minipage}{.8\linewidth}
    \vspace{1cm}
    \begin{center}
      {\small \textbf{Susanne Reffert}}
    \end{center}
    \vspace{1cm}
    \begin{minipage}{\linewidth}\centering
      {\itshape \footnotesize 
        Institute for the Physics and Mathematics of
        the Universe, \\
        The University of Tokyo, Kashiwa-no-Ha 5-1-5, \\
        Kashiwa-shi, 277-8568 Chiba, Japan.\\
      }
    \end{minipage}
  \end{minipage}
}

\date{}

\begin{document}

\begin{titlepage}
  \maketitle
  \thispagestyle{empty}

 \vfill

\abstract{\RaggedLeft In this note, an important extension to the recent construction of the fluxtrap background is presented. The fluxtrap is a closed string background based on the Melvin solution corresponding to the Omega deformation of flat space.  In this note, we introduce the mechanisms to extend it from $\varepsilon_1=-\varepsilon_2\in \mathbb{R}$ to more general values of $\varepsilon_1$ and $\varepsilon_2$ in $\mathbb{C}$.}
\vspace{2cm}
\end{titlepage}

\clearpage

\section{Introduction}
\label{sec:refin-compl-omega}

The $\Omega$--background was first introduced in \cite{Moore:1997dj, Lossev:1997bz} and used in \cite{Nekrasov:2002qd} as a regulator for the calculation of the volume of the instanton moduli space and has seen many applications since. Most recently, it has appeared in the context of the gauge/Bethe correspondence~\mbox{\cite{Nekrasov:2009rc, Nekrasov:2010ka}} and in the context of topological string theory \cite{Hollowood:2003cv, Iqbal:2007ii, Antoniadis:2010iq, Krefl:2010fm,Huang:2010kf, Aganagic:2011mi}. We won't have anything to say about the topological string interpretation of the $\varepsilon$--parameters of the $\Omega$--deformation, but will examine the subject from the point of view of the \emph{physical} string theory.

In~\cite{Hellerman:2011mv}, a certain closed string background based on the Melvin solution, the so-called \emph{fluxtrap}, was shown to be the string theory realization of the $\Omega$--background in the special case where the deformation parameters $\varepsilon_1$, $\varepsilon_2$ take the values $\varepsilon_1=-\varepsilon_2=\varepsilon\in \mathbb{R}$. 
This \emph{real fluxtrap} solution is constructed starting from a fluxbrane or Melvin solution in type IIB string theory. Here, an identification with parameter $m=\varepsilon\in \mathbb{R}$ is imposed on the angular variable of the $(45)$--plane, and another with parameter $-m=-\varepsilon$ on the angular variable in the $(67)$ plane. A T--duality in the $x_8$ direction eliminates degrees of freedom which are incompatible with the identifications and results in the fluxtrap solution in type IIA which has a manifestly non-flat metric and a B--field and preserves 16 real supercharges.

This background is of special interest since it is possible to construct a brane realization of the gauge/Bethe correspondence in it \cite{Orlando:2010uu, Orlando:2010aj, Hellerman:2011mv}.  This case, however, is not the most natural for most other applications. In most cases, a complex $\varepsilon$ is called for, or two $\varepsilon$--parameters which are independent of each other. It is therefore important to answer the question of how these generalizations can be constructed within string theory. In the present note, we therefore extend the fluxbrane construction to cover  also $\varepsilon\in \mathbb{C}$ and $\varepsilon_1+\varepsilon_2\neq0$\footnote{As we will see, the combination $\varepsilon_1,\,\varepsilon_2\in \mathbb{C},\ \varepsilon_1+\varepsilon_2\neq0$ cannot be combined with all brane setups.}. We will see in the following that the mechanisms for the two types of generalizations are fundamentally different from each other.

\bigskip
When D2--branes suspended between parallel NS5--branes are inserted in a compatible way, the bulk background discussed in~\cite{Hellerman:2011mv} corresponds to turning on a real component of the \emph{twisted mass} term of the adjoint field in the low energy effective gauge theory living on the D2--brane. In terms of this mass term, a complex $\varepsilon$ corresponds to a complex twisted mass term. To turn on such a complex mass term, we find that we must introduce two independent sets of identifications related to two T--dualities.

In order to add a second,  independent $\varepsilon$ parameter, we find on the other hand that the discrete identifications of the original fluxbrane solution must run over more coordinate directions. As we will see, having two independent $\varepsilon$ parameters is incompatible with the presence of both NS5--branes and D2--branes, but in the 3-dimensional gauge theory describing extended D2--branes, they correspond to \emph{real mass terms} for the three complex scalar fields in the adjoint chiral multiplets.

In the following, the conventions and coordinate directions are taken to be the same as in~\cite{Hellerman:2011mv}~and are
summarized in Table~\ref{tab:NS5-embedding}. The $\varepsilon$--parameters of the bulk appear as different parameters in the gauge theory (such as the twisted mass or the gauge coupling) depending on the chosen brane embedding. In order to realize an $\Omega$--deformed theory such as \emph{e.g.} in~\cite{Nekrasov:2003rj}, a Euclidean brane must be inserted as in Table~\ref{tab:NS5-embedding}. To realize on the other hand the setup necessary for the gauge/Bethe correspondence\footnote{Here, we will confine ourselves only to the simplest case, namely the one corresponding to the XXX$_{1/2}$ spin chain with periodic boundary conditions and without inhomogeneities.}, we need a stack of D2--branes suspended between two parallel NS5--branes and a stack of D4--branes as given in Table~\ref{tab:NS5-embedding}. It turns out that only the real and complex fluxtrap backgrounds with $\varepsilon_1+\varepsilon_2=0$ are compatible with the full gauge/Bethe setup.

\begin{table}
  \centering
  \begin{tabular}{lcccccccccc}
    \toprule
    direction & 0 & 1 & 2 & 3 & 4 & 5 & 6 & 7 & 8 & 9 \\  \midrule 
    fluxtrap & $\times $ & $\times$ & $\times $ & $\times $ & & & & & $\circ$ & $\circ$ \\ \midrule 
    NS5 & $\times $ & $\times $ & & & & & $\times $ & $\times $ & $\times $ & $\times$\\ 
    D2 & $\times $ & $\times $ & $\times $ \\ 
    D4 & $\times $ & $\times $ & & $\times $ & $\times $ & $\times $ \\ \midrule
    E4 & & & & & $\times $ & $\times $ & $\times $ & $\times $  \\
    \bottomrule
  \end{tabular}
  \caption{Embedding of the D2--brane with respect to the
    NS5 fluxtrap. The $\circ$ in the fluxtrap marks the directions of the T--dualities.}
  \label{tab:NS5-embedding}
\end{table}

\bigskip

As already discussed, there are two natural ways to generalize the fluxtrap background:
\begin{enumerate}
\item \emph{Adding an imaginary component to the two-dimensional twisted mass.} This
  is obtained by taking also $x_9$ to be periodic and generalizing the
  fluxbrane to a double identification in $x_8$ and $x_9$. As we will see, this preserves the same amount of supersymmetry as the real case.
\item \emph{Using two independent parameters $\varepsilon_1,\ \varepsilon_2$ for the identifications in
  the $(45)$ and $(67)$ planes.} In order to preserve supersymmetry it
  is necessary to introduce a third identification, which can be:
  \begin{itemize}
  \item in the $(39)$ plane for the theory on D2--branes without NS5;
  \item in the $(23)$ plane for the theory on E4--branes in presence of NS5.
  \end{itemize}
  This procedure preserves only half as many supercharges as the one above.
\end{enumerate}
We will explore the first possibility in Section~\ref{sec:complex-omega-background}, and the second in Section~\ref{sec:refined}.

\section{The complex $\Omega$--background}
\label{sec:complex-omega-background}

In the context of two dimensional gauge theories, the identification parameter of the fluxtrap background translates into twisted masses of the adjoint and bifundamental fields encoded by the fluctuations of the D2--branes set in this background. In general, these twisted masses are complex parameters. It is therefore vital to produce both components of the twisted mass in order to capture the most general case. In~\cite{Hellerman:2011mv}, it was shown how imposing identifications with a real parameter $m_8$  linked to a periodic variable $\tilde u$ in which then a T--duality is performed leads to a real twisted mass. How now does one produce the second component of a complex twisted mass term? In order to produce a second real component, it is necessary to introduce a second periodic variable $\tilde v$ linked to a second set of identifications with parameter $m_9$, and to perform a second T--duality in direction $\tilde v$. This makes sense since the complex combination $\tilde u+i\tilde v$ corresponds to the the complex scalar of the twisted chiral multiplet in the gauge theory.

\subsection{Complex fluxbrane and fluxtrap}
\label{sec:fluxbrane}
The complex $\Omega$--background, \emph{i.e.} $\varepsilon_1=-\varepsilon_2=\varepsilon\in \mathbb{C}$, is obtained by starting with either the flat
space or the NS5 solution, imposing the fluxbrane identifications and performing T--dualities in the
directions $x_8$ and $x_9$. Note that these directions are both parallel to the
NS5 and do not turn it into a Taub--NUT space. We can treat the cases with and without
NS5--branes in parallel since the flat case corresponds to $N_5 = 0 $. Some
extra care is needed in the definition of the Killing spinors, because
the flat case preserves $16$~supercharges while the background with NS5--branes preserves only $8$.

Let us start from the standard NS5 solution in type IIA string theory with the fields
\begin{align}
  {\di s}^2 &= - \di x_0^2 + \di x_1^2 + U \left[ \di x_2^2 + \di x_3^2 + \di \rho_1^2 +
    \rho_1^2 \di \theta_1^2 \right] + \di \rho_2^2 + \rho_2^2 \di
  \theta_2^2 +  \di \wt x_8^2 + \di \wt x_9^2 \, , \\
  B &= U_{,3} \left( - \left( x_3^2 + \rho_1^2 \right) \di x_2 + x_2
    x_3 \di x_3 + x_2 \rho_1 \di \rho_1 \right) \wedge
  \di \theta_1 \, , \\
  \Phi &= \frac{1}{2}\log U \, ,
\end{align}
where
\begin{align}
  U &= 1 + \frac{N_5\, \alpha'}{x_2^2 + x_3^2 + \rho_1^2} \, ; &   U_{,3} = \frac{\di }{\di x_3} U = - \frac{2 N_5\, \alpha' x_3 }{\left( x_2^2 + x_3^2 + \rho_1^2 \right)^2} \, .
\end{align}
Let $\wt x_8 $ and $\wt x_9 $ be periodic variables with periods $2 \pi \wt R_8 $ and $2 \pi \wt R_9 $ and introduce two $2 \pi$--periodic variables $\wt u $ and $\wt v$:
\begin{align}
  \wt x_8 &= \wt R_8\, \wt u \, ,& \wt x_9 &= \wt R_9\, v \, .
\end{align}
We can impose the two \emph{independent} sets of identifications
\begin{align}
    \begin{cases}
    \wt u \simeq \wt u + 2 \pi \, k_1 \, , \\
    \theta_1 \simeq \theta_1 + 2 \pi\, m_8 \wt R_8 \, k_1 \, , \\
    \theta_2 \simeq \theta_2 - 2 \pi\, m_8 \wt R_8 \, k_1 \, ,
  \end{cases} &&
    \begin{cases}
    \wt v \simeq \wt v + 2 \pi \, k_2 \, , \\
    \theta_1 \simeq \theta_1 + 2 \pi\, m_9 \wt R_9 \, k_2 \, , \\
    \theta_2 \simeq \theta_2 - 2 \pi\, m_9 \wt R_9 \, k_2 \, ,
  \end{cases}
\end{align}
where $k_1,\,k_2\in \mathbb{Z}$.
It is convenient to introduce the $2 \pi$--periodic variables $\phi_1 $ and $\phi_2 $ defined by
\begin{equation}
  \begin{cases}
    \phi_1 = \theta_1 - m_8 \wt R_8\, \wt u - m_9 \wt R_9\, \wt v \, ,\\
    \phi_2 = \theta_2 + m_8 \wt R_8\, \wt u + m_9 \wt R_9\, \wt v   \, ,
  \end{cases}
\end{equation}
and rewrite the bulk fields in the form
\begin{align}
  \begin{split}
    & {\di s}^2 = - \di x_0^2 +\di x_1^2  + U \left[ \di x_2^2 +
      \di x_3^2 + \di \rho_1^2 + \rho_1^2 \left( \di \phi_1 + m_8 \wt
        R_8 \di \wt u + m_9 \wt R_9 \di \wt v \right)^2 \right]  \\
    & \hspace{3em}+ \di \rho_2^2 + \rho_2^2 \di \left( \di \phi_2 - m_8 \wt R_8 \di
      \wt u - m_9 \wt R_9 \di \wt v \right)^2 + \wt R_8^2 \di \wt u+ \wt R_9 \di \wt v^2 \, ,
  \end{split} \\
   & B= U_{,3} \left( - \left( x_3^2 + \rho_1^2 \right) \di x_2 + x_2
    x_3 \di x_3 + x_2 \rho_1 \di \rho_1 \right) \wedge
  \left( \di \phi_1 + m_8 \wt
    R_8 \di \wt u + m_9 \wt R_9 \di \wt v \right) \, ,\\
  & \Phi = \Phi_0 +  \frac{1}{2}\log U \, .
\end{align}
These expressions become more transparent if we introduce the new variables
\begin{align}
  z_1 = x_4 + \imath x_5 &= \rho_1 \eu^{\imath \phi_1}\,, & z_2 = x_6 + \imath x_7 &=
  \rho_2 \eu^{\imath  \phi_2}\,, & \wt \zeta &= \wt x_8 + \imath \wt x_9 \, .
\end{align}
Now the metric becomes
\begin{multline}
  \di s^2 = -\di x_0^2 +\di x_1^2 + U \left[ \di x_2^2 + \di x_3^2 + \sum_{i=4}^5
    \left( \di x_i^2 + \varepsilon V^i \di \wt \zeta + \bar
      \varepsilon V^i \di \bar{ \wt \zeta} \right)^2 \right] + \\
 + \sum_{i=6}^7 \left( \di x_i^2 + \varepsilon V^i \di \wt \zeta + \bar
    \varepsilon V^i \di \bar {\wt \zeta} \right)^2 + \di \wt \zeta \di \bar{\wt \zeta}  \, ,
\end{multline}
where 
\begin{equation}
    V^i \del_i = - x^5 \del_{4} + x^4 \del_{5} + x^7 \del_{6} -
  x^6 \del_{7} = \del_{\phi_1} - \del_{\phi_2} \, ,  
\end{equation}
and $\varepsilon$ is the complex parameter
\begin{equation}
  \varepsilon = \frac{ m_8 - i m_9 }{2} \, .
\end{equation}
This reproduces exactly the expression given for example in~\cite{Nekrasov:2003rj}. In the usual notation this corresponds to
\begin{equation}
  \varepsilon = \varepsilon_1 = -\varepsilon_2 \, .  
\end{equation}
This is what we will call the \emph{complex fluxbrane}.

At this point, starting from the expression in cylindrical coordinates,
we can T--dualize in $\wt u $ and $\wt v$ to obtain the \emph{complex
fluxtrap} background, again in type IIA.
In the absence of NS5--branes, when $U = 1$, the fields in the bulk are given by
\begin{align}
  \begin{split}
    \di s^2 =& \di \vec x^2_{0123} + \di \rho_1^2 + \rho_1^2 \di
    \phi_1^2 + \di \rho_2^2 + \rho_2^2 \di \phi_2^2  \\
    & + \frac{1}{\Delta^2} \left( R_8^2 \di u^2 + R_9^2 \di v^2 + \left(
        \rho_1^2 + \rho_2^2 \right) \left( R_8 m_9 \di v - R_9 m_8 \di
        u \right)^2\right.  \\ 
        &\left.-\left( m_8^2 + m_9^2 \right) \left( \rho_1^2 \di
        \phi_1- \rho_2^2 \di \phi_2 \right)^2 \right),
  \end{split} \\
   B =& \frac{1}{\Delta^2} \left( \rho_1^2 \di \phi_1 - \rho_2^2 \di
    \phi_2 \right) \wedge \left( m_8 R_8 \di u + m_9 R_9 \di v \right)
  \, ,  \\
   \eu^{-\Phi} =& \eu^{-\Phi_0} \frac{\alpha' \Delta}{R_8 R_9} \, ,
\end{align}
where $u $ and $v$ are the T--dual variables to $\wt u $ and $\wt v$, 
\begin{align}
  R_8 &= \frac{\alpha'}{\wt R_8} \, ,& R_9 &= \frac{\alpha'}{\wt R_9 } \, ,
\end{align}
and
\begin{equation}
  \Delta^2 = 1 + \left( m_8^2 + m_9^2 \right) \left( \rho_1^2 +
    \rho_2^2 \right) \, .
\end{equation}
In rectilinear coordinates, $\zeta = R_8 u + \imath R_9 v $,
\begin{align}
\begin{split}
   \di s^2 =& \di \vec x^2_{01234567} 
    + \frac{1}{\Delta^2} \left( \di \zeta \di \bar \zeta - \left(
       z_1 \bar z_1 + z_2 \bar z_2 \right) \left( \varepsilon \di \bar
       \zeta - \bar \varepsilon \di \zeta \right)^2\right. \\
       &\left.- \bar
     \varepsilon \varepsilon \left( z_1 \di \bar z_1 + \bar z_1 \di
       z_1 - z_2 \di \bar z_2 - \bar z_2 \di z_2 \right)^2 \right), \end{split} \\
   B =& \frac{\imath}{2\Delta^2} \left( z_1 \di \bar z_1 + \bar z_1
     \di z_1 - z_2 \di \bar z_2 - \bar z_2 \di z_2
   \right) \wedge \left(  \varepsilon \di \bar
       \zeta - \bar \varepsilon \di \zeta \right)
  \, ,  \\
  \eu^{-\Phi} =& \eu^{-\Phi_0} \frac{\alpha' \Delta}{R_8 R_9} \, ,
\end{align}
and
\begin{equation}
  \Delta^2 = 1 + 4\,  \varepsilon \bar \varepsilon \left( z_1 \bar z_1 +
  z_2 \bar z_2 \right) \, .
\end{equation}
All the unphysical degrees of freedom have been eliminated by the T--dualities and we are left with a non-flat metric and a non-zero B--field.

Even though it is not a priori clear that this should work, it is possible to add NS5--branes to the fluxbrane background in a natural way. In the presence of an NS5--brane, the calculation works analogously, but the expressions for the bulk fields after the two T--dualities become more cumbersome. Instead of the metric, we will write down the inverse vielbein which contains the same information ($g^{\mu\nu}=e_a^\mu e_b^\nu\eta^{ab}$) but takes a simpler form:
\begin{align}
e_0&=\del_0,\\
e_1&=\del_1,\\
e_2&=\frac{1}{\sqrt U}\del_2+\frac{U_{,3}(x_3^2+\rho_1^2)}{\sqrt U}\left(\frac{m_8}{R_8}\del_u+\frac{m_9}{R_9}\del_v\right),\\
e_3&=\frac{1}{\sqrt U}\del_3-\frac{U_{,3}x_2x_3}{\sqrt U}\left(\frac{m_8}{R_8}\del_u+\frac{m_9}{R_9}\del_v\right),\\
e_4&=\frac{1}{\sqrt U}\del_{\rho_1}-\frac{U_{,3}x_2\,\rho_1}{\sqrt U}\left(\frac{m_8}{R_8}\del_u+\frac{m_9}{R_9}\del_v\right),\\
e_5&=\frac{1}{\sqrt U \rho_1}\del_{\phi_1}+\sqrt U\, \rho_1\left(\frac{m_8}{R_8}\del_u+\frac{m_9}{R_9}\del_v\right),\\
e_6&=\del_{\rho_2},\\
e_7&=\frac{1}{\rho_2}\del_{\phi_2}-\rho_2\left(\frac{m_8}{R_8}\del_u+\frac{m_9}{R_9}\del_v\right),\\
e_8&=-m_8\del_{\phi_1}+m_8\del_{\phi_2}+\frac{1}{R_8}\del_u,\\
e_9&=-m_9\del_{\phi_1}+m_9\del_{\phi_2}+\frac{1}{R_9}\del_v.
\end{align}
The B--field is given by
\begin{multline}
B=\frac{1}{\Delta^2}\left(U_{,3}[(1+(m_8^2+m_9^2)\rho_2^2)\di\phi_1+(m_8^2+m_9^2)\rho_2^2\di\phi_2]\wedge\right.\\ 
[-(x_3^2+\rho_1^2)\di x_2+x_2x_3\di x_3+x_2\rho_1 \di \rho_1]\\
\left.+(U\,\rho_1^2\di\phi_1-\rho_2^2\di\phi_2)\wedge(m_8R_8\,\di u+m_9R_9\di v)\right),
\end{multline}
where $\Delta$ is now given by $\Delta=1+(m_8^2+m_9^2)(U\,\rho_1^2+\rho_2^2)$,
and the dilaton by
\begin{equation}
  \eu^{-\Phi} = \eu^{-\Phi_0} \frac{\alpha'\Delta}{R_8 R_9\sqrt U},
\end{equation}
where $\Phi_0$ fixes the gauge coupling of the effective theory living on the branes which can live in this background.

\subsection{Supersymmetries for the complex fluxtrap}
\label{sec:supersymmetry}

The analysis of the supersymmetries follows the same pattern as the one for the real
$\Omega$--background discussed in~\cite{Hellerman:2011mv}.
The original 32 real components of the Killing spinor are halved by the projector $\proj{NS5}_\pm=\tfrac{1}{2} \left( \Id \pm \Gamma_{2345}
  \right)$ due to the presence of the NS5--brane. Due to the fluxbrane background, only components compatible with the identifications remain preserved. The incompatible ones are projected out by $\proj{flux}_\pm = \tfrac{1}{2} \left( \Id \pm \Gamma_{4567} \right) $, reducing the supersymmetry by another one half. It is possible to relate the Killing spinors before and after T--duality. The transformation is in general non-local. It is possible to choose a suitable vielbein that greatly simplifies the expressions if we require the combination \( e^a_\mu \del X^\mu \) to remain invariant. In this case the left-moving Killing spinors remain invariant and the right-moving spinors are multiplied by an appropriate \( \Gamma \) matrix.
Making use of the T--dual vielbein (see~\cite{Hellerman:2011mv}), the Killing spinors in type IIA after two T--dualities are given by $K^{IIA}=\epsilon_L+\epsilon_R$ with
\begin{equation}
  \epsilon_L = \left( \Id  + \Gamma_{11} \right)\proj{flux}_- \proj{NS5}_- \exp [ \frac{\phi_1}{2} \Gamma_{45} ] \exp[ \frac{\phi_2}{2} \Gamma_{67} ]   \epsilon_0    
\end{equation}
and
\begin{equation}
  \epsilon_R =  \left( \Id - \Gamma_{11} \right)\Gamma_v \Gamma_u \proj{flux}_- \proj{NS5}_+ \exp [ \frac{\phi_1}{2} \Gamma_{45} ] \exp[ \frac{\phi_2}{2} \Gamma_{67} ]  \epsilon_1  \,,
\end{equation}
where \(\epsilon_0 \) and \(\epsilon_1 \) are constant Majorana
spinors and \( \Gamma_u \) and \( \Gamma_v \) are the gamma matrices
in the direction of the T--dualities, normalized to \( (\Gamma_u)^2 = (
\Gamma_v )^2 = 1 \). It is important to realize that with this normalization, the Gamma
matrix in the direction of the T--duality does not change under this
T--duality (it does, however, change under a T--duality in a different direction).  Each of the projectors $\proj{flux}_\pm$, $\proj{NS5}_\pm$, and $\left( \Id \pm \Gamma_{11} \right)$ reduces the supersymmetry by one half. 

We now need to show that the order of the two T--dualities does not matter for the counting.
We will argue that the following diagram commutes:
\tikzset{node distance=2cm, auto}

\begin{equation}
\begin{tikzpicture}
\node (1) {$(1)$};
\node (2) [below of=1] {$(2)$};
\node (2p) [right of=1] {$(2')$};
\node (3) [below of=2p] {$(3) \equiv (3')$};
\draw[->] (1) to node {$u$} (2);
\draw[->] (1) to node {$v$} (2p);
\draw[->] (2) to node {$v$} (3);
\draw[->] (2p) to node {$u$} (3);
\end{tikzpicture}
\end{equation}
Here, \( (1) \) is the initial fluxbrane, \( (2) \) is the
configuration after T--duality in \(u \), \( (3) \) is the
configuration after double T--duality and \( (2') \) the configuration
after one T--duality in \( v \).

The above diagram corresponds to requiring the T--duality
operator to act as an \emph{intertwiner}, \emph{i.e.}
\begin{equation}\label{eq:inter}
   \Gamma_v^{(3)} \Gamma_u^{(2)} = - \Gamma_u^{(3)} \Gamma_v^{(2')}  ,
\end{equation}
The $\Gamma_{u,v}^{(i)}$ are given by 
\begin{equation}
\Gamma_u^{(i)}=\tfrac{1}{({g_{uu}^{(i)}})^{1/2}}e\ud{(i)m}{u}\Gamma_m.
\end{equation}
It is clear that  it is essential to evaluate the $\Gamma_{u,v}^{(i)}$ on the right vielbein. Under T--duality in $u$, the vielbein transforms as follows (see also~\cite{Hellerman:2011mv}):
\begin{equation}
\label{eq:vielbein-T--duality}
  \begin{dcases}
    e\ud{(2)m}{u}=\frac{\alpha'}{g^{(1)}_{uu}}e\ud{(1)m}{u}\,, \\
    e\ud{(2)m}{v} = e\ud{(1)m}{v} -
    \frac{g^{(1)}_{vu}+ B^{(1)}_{vu}}{g^{(1)}_{uu}} e\ud{(1)m}{u} .
  \end{dcases}
\end{equation}
With the above remark, we see that $\Gamma_v^{(3)}=\Gamma_v^{(2)}$ and $\Gamma_u^{(3)}=\Gamma_u^{(2')}$.
The condition (\ref{eq:inter}) now becomes
\begin{equation}
\{\Gamma_v^{(2)}\Gamma_u^{(2)},\Gamma_u^{(2')}\Gamma_v^{(2')} \}=0,
\end{equation}
which can be verified explicitly in a straightforward way with the help of the usual T--duality rules.

To conclude, depending on whether or not we include NS5--branes, we are left with 8, resp. 16 real supercharges, the same amount as in the real fluxtrap.

\section{The $\Omega$--background with $\varepsilon_1 + \varepsilon_2 \neq 0$}\label{sec:refined}

The $\Omega$--background as discussed in~\cite{Moore:1997dj, Lossev:1997bz} comes with two independent deformation parameters $\varepsilon_1,\, \varepsilon_2$. In order to make this general $\Omega$--background and its applications accessible via string theory, it is essential to formulate a fluxtrap background with $\varepsilon_1 + \varepsilon_2 \neq 0$.

While we have seen in the last section that in order to produce the two real components of a complex deformation parameter we need to perform two T--dualities along two periodic variables, the approach for producing two independent real deformation parameters works differently, namely by introducing identifications in a third coordinate plane linked to a single periodic variable.
We will refer to this background as the \emph{refined fluxtrap}, in reference to the refined topological string where the second $\varepsilon$ acts as a refinement parameter.

\subsection{The real refined fluxtrap}
We will see that in order to have $\varepsilon_1 + \varepsilon_2 \neq 0$ while still preserving
part of the supersymmetry, we need to introduce another
identification on top of those in the $(45)$ and $(67)$ planes to compensate their actions (see Sec.~\ref{sec:supersymm-refin-flux}). 

Let us consider the case of the bulk background with $\varepsilon_1,\ \varepsilon_2\in \mathbb{R}$. Since only one T--duality is needed, we start out in type IIB string theory this time. The third
identification can be chosen to happen in the $(x_3, x_9)$ plane which still allows us to place a D2--brane in the $012$ directions. It is, however, not possible to also add an NS5--brane.\footnote{We could choose instead to do the third identification in the (23) direction, which would enable us to have an NS5--brane as in Table~\ref{tab:NS5-embedding}, but preclude us from adding also the D2.}
We start out with the flat metric
\begin{equation}
\di s^2=\di \vec x^2_{012} + \di \wt x_8^2+\di \rho_1^2 + \rho_1^2 \di
    \theta_1^2 + \di \rho_2^2 + \rho_2^2 \di \theta_2^2+\di \rho_3^2 + \rho_3^2 \di \theta_3^2.
\end{equation}
We now impose three sets of identifications with two independent parameters,
\begin{align}
  \begin{cases}
    \wt u \simeq \wt u + 2 \pi \, k_1 \, , \\
    \theta_1 \simeq \theta_1 + 2 \pi\, m_{81} \wt R_8 \, k_1 \, , \\
    \theta_2 \simeq \theta_2 + 2 \pi\, m_{82} \wt R_8 \, k_1 \, , \\
    \theta_3 \simeq \theta_3 - 2 \pi \left( m_{81} + m_{82} \right)
    \wt R_8 k_1 \, .
  \end{cases} 
\end{align}
The need for the identification in a third plane with parameter $-(m_{81} + m_{82})$ will become clear in the discussion of the supersymmetry properties of the refined fluxbrane background in Section~\ref{sec:supersymm-refin-flux}.

With the usual procedure, we can introduce the $2\pi$--periodic
variables $\phi_1, \phi_2, \phi_3$:
\begin{equation}
  \begin{cases}
    \phi_1 = \theta_1 - m_{81} \wt R_8\, \wt u \, , \\
    \phi_2 = \theta_2 - m_{82} \wt R_8\, \wt u \, , \\
    \phi_3 = \theta_3 + \left( m_{81} + m_{82} \right) \wt R_8\, \wt u \, . \\
  \end{cases}
\end{equation}
It is convenient to introduce the coordinates
\begin{align}
  \wt x_8 &= \wt R_8\, \wt u \, , & z_1 &= x_4 +
  \imath x_5 = \rho_1 \eu^{\imath \phi_1} \, , & z_2 &= x_6 + \imath x_7 =
  \rho_2 \eu^{\imath \phi_2} \, , & z_3 &= x_3 + \imath x_9 = \rho_3
  \eu^{\imath \phi_3} \, ,
\end{align}
and the two real parameters $\varepsilon_1 = m_{81},\  \varepsilon_2 =m_{82}$
to write the (flat) metric in the form
\begin{multline}
  \di s^2 = - \di x_0^2 + \di x_2^2 + \di \wt x_8^2  
  + \sum_{i=4}^5 \left( \di x_i^2 + \varepsilon_1 V^i \di \wt x_8  \right)^2
  + \sum_{i=6}^7 \left( \di x_i^2 + \varepsilon_2 V^i \di \wt x_8  \right)^2 \\
  + \sum_{i=\set{3,9}} \left( \di x_i^2 -
    \left( \varepsilon_1 + \varepsilon_2 \right) V^i \di \wt x_8  \right)^2 ,
\end{multline}
where
\begin{equation}
  V^i \del_i = - x^5 \del_{4} + x^4 \del_{5} - x^7 \del_{6} +
  x^6 \del_{7}  -  x^3 \del_9 + x^9 \del_3 = \del_{\phi_1} + \del_{\phi_2} + \del_{\phi_3} \, . 
\end{equation}
After T--duality, we have the following bulk fields:
\begin{align}
  \begin{split}
    \di s^2 =& \di \vec x^2_{012} + \di \rho_1^2 + \di \rho_2^2 + \di \rho_3^2  + \frac{1}{\widetilde\Delta^2} \left[ R_8^2 \di u^2 + \left(1+\varepsilon_2^2\rho_2^2+(\varepsilon_1+\varepsilon_2)^2\rho_3^2\right)\rho_1^2 \di \phi_1^2\right.\\
    & + \left(
        1+\varepsilon_1^2\rho_1^2+(\varepsilon_1+\varepsilon_2)^2\rho_3^2 \right)\rho_2^2\di\phi_2^2 +\left( 1+\varepsilon_1^2\rho_1^2+\varepsilon_2^2\rho_2^2 \right)\rho_3^2\di\phi_3^2\\
        &\left.+2\,\varepsilon_1\rho_1^2\left(-\varepsilon_2\rho_2^2\di\phi_1\di\phi_2+(\varepsilon_1+\varepsilon_2)\rho_3^2\di\phi_1\di\phi_3\right)+2\,\varepsilon_2(\varepsilon_1+\varepsilon_2)\rho_2^2\rho_3^2\di\phi_2\di\phi_3\right],
  \end{split} \\
   B =& \frac{R_8}{\widetilde\Delta^2} \left( -\varepsilon_1\rho_1^2\di \rho_1\wedge\di v-\varepsilon_2\rho_2^2\di\rho_2\wedge\di v+(\varepsilon_1+\varepsilon_2)\rho_3^2\di u\wedge \di v \right),  \\
  \eu^{-\Phi} =& \eu^{-\Phi_0} \frac{\sqrt\alpha' \widetilde\Delta}{R_8} \, ,
\end{align}
 where 
 \begin{equation}
 \widetilde\Delta^2=1+\varepsilon_1^2\,\rho_1^2+\varepsilon_2^2\,\rho_2^2+(\varepsilon_1+\varepsilon_2)^2\rho_3^2.
 \end{equation}
 We refer to this solution as the \emph{real refined fluxtrap}. 

\subsection{The complex refined fluxtrap}

Of course, we can combine the two mechanisms introduced above and construct a fluxtrap background with $\varepsilon_1 + \varepsilon_2 \neq 0$, where now $\varepsilon_{1,2}\in\mathbb{C}$.  Since this requires a second T--duality in the $v$ or $x_9$--direction, the extra identification cannot take place in the $(x_3,\ x_9)$--coordinates but must be relegated to the $(x_2,\ x_3)$--directions. Since now, $x_2$ is not free anymore, this precludes us from placing a D2--brane in the $012$--directions, but still leaves us the option of inserting an NS5 or an E4--plane as indicated in Table~\ref{tab:NS5-embedding}.
In flat space, the \emph{complex refined} fluxbrane is therefore introduced as follows.
We start out from the flat metric
\begin{equation}\label{eq:refcpxmet}
\di s^2=\di \vec x^2_{01} + \di \wt x_8^2+\di \wt x_9^2+\di \rho_1^2 + \rho_1^2 \di
    \theta_1^2 + \di \rho_2^2 + \rho_2^2 \di \theta_2^2+\di \rho_3^2 + \rho_3^2 \di \theta_3^2.
\end{equation}
Impose the identifications
\begin{align}
  \begin{cases}
    \wt u \simeq \wt u + 2 \pi \, k_1 \, , \\
    \theta_1 \simeq \theta_1 + 2 \pi\, m_{81} \wt R_8 \, k_1 \, , \\
    \theta_2 \simeq \theta_2 + 2 \pi\, m_{82} \wt R_8 \, k_1 \, , \\
    \theta_3 \simeq \theta_3 - 2 \pi \left( m_{81} + m_{82} \right)
    \wt R_8 k_1 \, ,
  \end{cases} &&
    \begin{cases}
      \wt v \simeq \wt v + 2 \pi \, k_2 \, , \\
      \theta_1 \simeq \theta_1 + 2 \pi\, m_{91} \wt R_9 \, k_2 \, , \\
      \theta_2 \simeq \theta_2 + 2 \pi\, m_{92} \wt R_9 \, k_2 \, , \\
      \theta_3 \simeq \theta_3 - 2 \pi \left( m_{91} + m_{92} \right)
      \wt R_9 k_2 \, .
  \end{cases}
\end{align}
With the usual procedure, we can introduce the $2\pi$--periodic
variables $\phi_1, \phi_2, \phi_3$:
\begin{equation}
  \begin{cases}
    \phi_1 = \theta_1 - m_{81} \wt R_8\, \wt u - m_{91} \wt R_9\, \wt v \, , \\
    \phi_2 = \theta_2 - m_{82} \wt R_8\, \wt u - m_{92} \wt R_9\, \wt v \, , \\
    \phi_3 = \theta_3 + \left( m_{81} + m_{82} \right) \wt R_8\, \wt u +
    \left( m_{91} + m_{92} \right) \wt R_9\, \wt v \, . \\
  \end{cases}
\end{equation}
It is convenient to introduce the coordinates
\begin{align}
  \wt \zeta &= \wt R_8 \wt u + \imath \wt R_9 \wt v \, , & z_1 &= x_4 +
  \imath x_5 = \rho_1 \eu^{\imath \phi_1} \, , & z_2 &= x_6 + \imath x_7 =
  \rho_2 \eu^{\imath \phi_2} \, , & z_3 &= x_2 + \imath x_3 = \rho_3
  \eu^{\imath \phi_3} \, ,
\end{align}
and the two complex parameters
\begin{align}
  \varepsilon_1 &= \frac{m_{81} - \imath m_{91}}{2} \, , & \varepsilon_2 &=
  \frac{m_{82} - \imath m_{92}}{2} \, ,
\end{align}
to write the (flat) metric in the form
\begin{multline}
  \di s^2 = - \di x_0^2 + \di x_1^2 + \di \wt \zeta \di \bar{\wt
    \zeta} \\
  + \sum_{i=4}^5 \left( \di x_i^2 + \varepsilon_1 V^i \di \wt \zeta +
    \bar \varepsilon_1 V^i \di \bar{ \wt \zeta} \right)^2
  + \sum_{i=6}^7 \left( \di x_i^2 + \varepsilon_2 V^i \di \wt \zeta +
    \bar \varepsilon_2 V^i \di \bar  {\wt \zeta} \right)^2 \\
  + \sum_{i=2}^3 \left( \di x_i^2 -
    \left( \varepsilon_1 + \varepsilon_2 \right) V^i \di \wt \zeta -
    \left( \bar \varepsilon_1 + \bar \varepsilon_2 \right) V^i \di
    \bar {\wt \zeta} \right)^2 ,
\end{multline}
where
\begin{equation}
  V^i \del_i \,= - x^5 \del_{4} + x^4 \del_{5} - x^7 \del_{6} +
  x^6 \del_{7}  -  x^3 \del_2 + x^2 \del_3 \,= \,\del_{\phi_1} + \del_{\phi_2} + \del_{\phi_3} \, . 
\end{equation}
The calculation of the fields after T--duality is cumbersome but straightforward. Due to their bulkiness, we refrain from giving the explicit expressions for metric and $B$--field\footnote{The author is happy to supply them upon direct inquiry.} and content ourselves with the dilaton:
\begin{equation}
\eu^{-\Phi} = \eu^{-\Phi_0} \frac{\alpha' \widetilde{\widetilde\Delta}}{R_8R_9},
\end{equation}
where now $\widetilde{\widetilde\Delta}^2$ takes the form
\begin{equation}\begin{split}
\widetilde{\widetilde\Delta}^2=&\,1+\left(m_{81}^2+m_{91}^2+(m_{81}m_{92}-m_{91}m_{82})^2\rho_2^2\right)\rho_1^2+\left(m_{82}^2+m_{92}^2\right)\rho_2^2\\
&+\left((m_{81}+m_{82})^2+m_{92}^2(1+m_{81}^2(\rho_1^2+\rho_2^2))-2\,m_{91}m_{92}(-1+m_{81}m_{82}(\rho_1^2+\rho_2^2))\right.\\
&\left.+\,m_{91}^2(1+m_{82}^2(\rho_1^2+\rho_2^2))\right)\rho_3^2\\
=&\,1+4\left|\epsilon_1\right|^2\rho_1^2+4\left|\epsilon_2\right|^2\rho_2^2+4\left|\epsilon_1+\epsilon_2\right|^2\rho_3^2-\left(\epsilon_1\overline\epsilon_2-\overline\epsilon_1\epsilon_2\right)\left(\rho_1^2\rho_2^2+\rho_1^2\rho_3^2+\rho_2^2\rho_3^2\right).
\end{split}\end{equation}
This solution we call the \emph{complex refined fluxtrap}.

\subsection{Supersymmetry of the complex refined fluxtrap}
\label{sec:supersymm-refin-flux}

The supersymmetries can be discussed directly in the complex case of the fluxbrane with three complex parameters in the
directions $(4,5),\ (6,7),\ (2,3)$ and be specialized to the real case if needed.
Our treatment is based on the one in~\cite{Russo:2001na}.
The Killing spinor for the metric~(\ref{eq:refcpxmet}) before the identifications has the form
\begin{equation}
  K = \eu^{\imath \theta_1\Gamma_{45}}e^{\imath
    \theta_2\Gamma_{67}}e^{\imath \theta_3\Gamma_{23}} \epsilon_0 \,,
\end{equation}
where $\epsilon_0$ is a Majorana spinor. After imposing the identifications, we use the coordinates
\begin{equation}
  \begin{cases}
    \theta_1 = \phi_1+\varepsilon_1\, {\wt \zeta} + \bar \varepsilon_1\,
    \bar{\wt \zeta},  \\
    \theta_2 = \phi_2 + \varepsilon_2\, {\wt \zeta} + \bar \varepsilon_2\,
    \bar {\wt \zeta}, \\
    \theta_3 = \phi_3 + \varepsilon_3\, {\wt \zeta} + \bar \varepsilon_3\,
    \bar{\wt \zeta} \, .
  \end{cases}
\end{equation}
The Killing spinor now takes the form
\begin{align}
  K = \eu^{\imath \phi_1\Gamma_{45}}e^{\imath
    \phi_2\Gamma_{67}}e^{\imath \phi_3\Gamma_{23}} e^{\imath \left(
      \varepsilon_1\Gamma_{45}+\varepsilon_2\Gamma_{67}+\varepsilon_3\Gamma_{23}
    \right) {\wt \zeta}+\text{c.c.}}  \epsilon_0 \, .
\end{align}
 In order for the Killing spinor to satisfy the boundary conditions of the fluxtrap, it has 
to be independent of ${\wt \zeta}$. We therefore need to impose
\begin{equation}
  \left( \varepsilon_1 \Gamma_{45} + \varepsilon_2 \Gamma_{67} +
    \varepsilon_3 \Gamma_{23} \right) \epsilon_0 = 0\,,
\end{equation}
or equivalently,
\begin{equation}
  \label{eq:cond}
  \varepsilon_1\Gamma_{45}\left[\mathbb{1}-\frac{\varepsilon_2}{\varepsilon_1}\Gamma_{4567}-\frac{\varepsilon_3}{\varepsilon_1}\Gamma_{2345}\right] \epsilon_0=0.
\end{equation}
This is fulfilled iff
\begin{equation}
  \label{eq:epp}
  \varepsilon_1+\varepsilon_2+\varepsilon_3=0\,,
\end{equation}
and
\begin{align}
  \left( \Gamma_{45}-\Gamma_{67}  \right) \epsilon_0=0,\label{eq:oneA}\\
  \left( \Gamma_{45}-\Gamma_{23} \right) \epsilon_0=0.\label{eq:twoA}
\end{align}
This can be easily seen as follows: with Eq.~(\ref{eq:epp}), the condition (\ref{eq:cond}) becomes
\begin{equation}
\left[(\mathbb{1}+\Gamma_{4567})+\frac{\varepsilon_3}{\varepsilon_1}(\Gamma_{4567}-\Gamma_{2345})\right]\epsilon_0=0.
\end{equation}
For this condition to hold for all $\varepsilon_3/\varepsilon_1$, we must impose
\begin{align}
  \left( \Id+\Gamma_{4567} \right) \epsilon_0&=0,\\
  \left( \Gamma_{4567}-\Gamma_{2345}\right)\epsilon_0&=0,
\end{align}
which is equivalent to 
\begin{align}
-\Gamma_{45}  \left( \Gamma_{45}-\Gamma_{67}  \right) \epsilon_0=0,\label{eq:oneB}\\
 \Gamma_{45} \left( \Gamma_{67}-\Gamma_{23} \right) \epsilon_0=0.\label{eq:twoB}
\end{align}
The Eq.~(\ref{eq:oneB}) corresponds directly to Eq.~(\ref{eq:oneA}), while subtracting Eq.~(\ref{eq:twoB}) from Eq.~(\ref{eq:oneB}) gives Eq.~(\ref{eq:twoA}).
With the above conditions, we can now rewrite the Killing spinors as
\begin{align}
  K = \eu^{\imath \phi_1\Gamma_{45}}e^{\imath
    \phi_2\Gamma_{67}}e^{\imath \phi_3\Gamma_{23}} (\Id-\Gamma_{4567})(\Id-\Gamma_{4523})  \epsilon_0 \, .
\end{align}
Each of the projectors reduces the supersymmetry by one half, which means that we are now left with 8 real supercharges.
To derive the Killing spinors after T--duality, the procedure is analogous to the one described in Sec.~\ref{sec:supersymmetry}.

The above treatment makes the need for introducing a third identification with parameter $\varepsilon_3$ manifest. If we require a supersymmetric setup with only two identifications with parameters $\varepsilon_1,\ \varepsilon_2$, we automatically arrive at the condition analogous to Eq.~(\ref{eq:epp}), $\varepsilon_1+ \varepsilon_2=0$, as discussed in~\cite{Hellerman:2011mv}. In order to arrive at a supersymmetric setup with $\varepsilon_1+ \varepsilon_2\neq0$, the introduction of a third identification such that the parameters fulfill $ \varepsilon_1+\varepsilon_2+\varepsilon_3=0$ is necessary.

\section{Summary and Outlook}

In this note, we have answered the important outstanding question of how to generalize the original fluxbrane construction given in~\cite{Hellerman:2011mv} to general values of the deformation parameters. The generalization to complex deformation parameters and the one to two independent deformation parameters are implemented in string theory via two conceptionally different mechanisms. 

In order to successfully combine a fluxtrap background with other elements such as D--, E-- or NS5-branes, we have to make sure that all $\del_{\theta_i}$ (corresponding to the complex planes $(\rho_i,\,\theta_i)$ in which the identifications happen) are Killing vectors. To ensure this, \emph{both} $\rho_i$ and $\theta_i$ need to be either transversal or parallel to the branes. While we have seen in this note that it is possible to combine D2-- and NS5--branes with a complex fluxtrap, we cannot fit an NS5 as needed for the gauge/Bethe setup into the refined real fluxtrap, or the D2s as needed into a complex refined fluxtrap background. On the other hand, it is possible to study an E4--brane as given in Table~\ref{tab:NS5-embedding} in the complex refined fluxtrap, even in conjunction with an NS5. Table~\ref{tab:summ} briefly summarizes the possibilities of interest here.\footnote{Note that this not an exhaustive list of all brane-like objects which could be studied inside a fluxtrap-type background to satisfy other interests.}
\begin{table}
  \centering
  \begin{tabular}{c c c }
    \toprule
    \ & $\varepsilon_1+ \varepsilon_2=0$ & $\varepsilon_1+ \varepsilon_2\neq0$ \\  \midrule 
    $\varepsilon_i\in \mathbb{R}$ & D2+NS5+D4; E4+NS5 & D2; NS5; E4+NS5 \\  \midrule 
    $\varepsilon_i\in \mathbb{C}$ & D2+NS5+D4; E4+NS5 &NS5; E4+NS5  \\ \bottomrule
  \end{tabular}
  \caption{Possible brane-like objects in various fluxtrap backgrounds}
  \label{tab:summ}
\end{table}
In order to preserve supersymmetry, we must satisfy $\sum_i\varepsilon_i=0$, where each independent $\varepsilon$ parameter reduces the conserved supercharges by one half.

To actually analyze the D--branes in the generalized backgrounds discussed here, a course analogous to the one described in detail in~\cite{Hellerman:2011mv} must be taken, but this goes beyond the purpose of this short note. To make an example, the low energy effective action at second order in the fields corresponding to a D2--brane in the complex fluxtrap will take the  form
\begin{equation}
  S \propto\int \di^3 \zeta \left[ -
    \dot X^\sigma \dot X_\sigma + (m_8^2+m_9^2) \left( \rho_1^2 + \rho_2^2
    \right) + \bar \psi\,
  \Gamma_0 \dot \psi + \frac{1}{2} \bar \psi \left( \Gamma_{45} -
    \Gamma_{67} \right)(m_8\Gamma_8+m_9\Gamma_9)  \psi \right] ,
\end{equation}
where the $X^\sigma$ are the bosonic coordinates in the transverse directions and the $\psi$ are the fermionic fields.

\bigskip
We are confident that the constructions shown here which live in the bulk of type II string theory will not only be of use in gaining further insights into the workings of the gauge/Bethe correspondence, but can also provide a different point of view for the study of questions in refined topological string theory.
\newpage

\subsection*{Acknowledgements}

The author is indebted to Simeon Hellerman and Domenico Orlando for numerous enlightening discussions and collaboration on related projects. The author furthermore acknowledges useful discussions with Kentaro Hori.
The author would like to thank the KITP in Santa Barbara and the SCGP in Stony Brook for hospitality. 
This research was supported in part by the National Science Foundation under Grant No. NSF PHY05-51164 and by the World Premier International Research
Center Initiative, MEXT, Japan.

\bibliography{RefinedNotes}

\end{document}